\documentclass[conference]{IEEEtran}
\IEEEoverridecommandlockouts
\usepackage{cite}
\usepackage{amsmath,amssymb,amsfonts}
\usepackage{algorithmic}
\usepackage{graphicx}
\usepackage{textcomp}
\usepackage{xcolor}
\usepackage{amsthm}
\usepackage{tabularx,booktabs}
\usepackage{eucal}
\usepackage{placeins}
\newcolumntype{C}{>{\centering\arraybackslash}X} 
\usepackage{lipsum}
\usepackage{stfloats}
\usepackage{siunitx}
\def\BibTeX{{\rm B\kern-.05em{\sc i\kern-.025em b}\kern-.08em
    T\kern-.1667em\lower.7ex\hbox{E}\kern-.125emX}}
    \newtheorem{theorem}{Theorem}
\newtheorem{lemma}[theorem]{Lemma}

\begin{document}

\title{Age of Information in a SWIPT and URLLC enabled Wireless Communications System
%\thanks{This work is funded  by ..}
}

	\author{\IEEEauthorblockN{Chathuranga M. Wijerathna Basnayaka\IEEEauthorrefmark{2}\IEEEauthorrefmark{1}, Dushantha Nalin K. Jayakody\IEEEauthorrefmark{3}
 \\Tharindu D. Ponnimbaduge Perera\IEEEauthorrefmark{2}, Timo T Hämäläinen\IEEEauthorrefmark{4} and Mário Marques da Silva\IEEEauthorrefmark{3} }

	\IEEEauthorblockA{\IEEEauthorrefmark{1}COPELABS, Lusófona University, Lisbon, PORTUGAL}
	\IEEEauthorblockA{\IEEEauthorrefmark{2}Centre for Telecommunications Research, School of Engineering,  Sri Lanka Technological Campus, Padukka, SRI LANKA}
	\IEEEauthorblockA{\IEEEauthorrefmark{3}Centro de Investiga\c {c}\"{a}o em Tecnologias - Aut\'{o}noma TechLab, Universidade Aut\'{o}noma de Lisboa, PORTUGAL}
	\IEEEauthorblockA{\IEEEauthorrefmark{4} Faculty of Information Technology, University of Jyväskylä, Jyväskylä, FINLAND}

		\IEEEauthorblockA{
		Email: {\{chathurangab, tharindupe\}@sltc.ac.lk , \{djayakody,mmsilva\}@autonoma.pt  } and timo.t.hamalainen@jyu.fi}
}
\maketitle

\begin{abstract}
This paper  estimates the freshness of the information in a wireless relay communication system that employs simultaneous wireless information and power transfer (SWIPT) operating under ultra-reliable low-latency communication (URLLC) constraints. The Age of Information (AoI) metric calculates the time difference between the current time and the timestamp of the most recent update received by the receiver is used here to estimate the freshness of information. The short packet communication scheme is used to fulfil the reliability and latency requirements of the proposed wireless network and its performance is analysed using finite block length theory. In addition, by utilising novel approximation approaches, expressions for the average AoI (AAoI) of the proposed system are derived. Finally, numerical analysis is used to evaluate and validate derived results.
\end{abstract}

\section{introduction}
The use cases of 5G communications are classified into three broad categories: enhanced Mobile BroadBand, URLLC  and mMTC. URLLC is necessary for mission-critical applications such as unmanned aerial vehicle (UAV) communication and process automation\cite{basnayaka2021agej,sharma2020communication,perera2020age}. The packet size should be extremely small in URLLC to facilitate low-latency transmission. The Shannon-Hartley Capacity theorem is no longer relevant in this scenario since the law of large numbers is invalid. However, using finite block length information theory, we can derive the achievable data rate under short packet communication as a function of the SNR, the block length and the decoding error probability\cite{polyanskiy2010channel}.  In addition, for these mission-critical applications, the freshness of the information is of high importance, along with the URLLC requirements. This has prompted growing interest in the age of information (AoI), a performance metric that quantifies the freshness of the information. \par On the other hand, SWIPT is an emerging technology for future wireless communication systems. In general, practical SWIPT receivers for energy harvesting (EH) and information decoding have been established through the use of power splitting (PS) and time switching (TS) techniques. Since EH shares the resources allocated to information transmission, this has an influence on the performance of wireless communication. However, minimal research has been conducted to analyse SWIPT-enabled relays using finite block length information theory and AoI. \par 
%Motivated by the features of SWIPT, URLLC, and cooperative communication and the need to measure the freshness of information in modern mission-critical applications,
This paper presents a wireless relay system with SWIPT for future mission-critical URLLC-enabled applications. To the best of the authors' knowledge, no prior study  has analysed AoI in a SWIPT and URLLC enabled wireless relay network. In this work, a two-way wireless relay system employs a nonlinear PS model for energy harvesting and short packet communication is employed to address the trade-off between reliability and latency. We derived an approximation for the average AoI (AAoI) of the proposed relay scheme under the finite blocklength constraint. In comparison to prior work on SWIPT, we examine the AoI performance of the proposed SWIPT system using threshold-based nonlinear EH. Furthermore, we examine the effect of various factors, including block length and packet size, on the weighed sum AAoI. 

\section{System Model}
\begin{figure}[!htbp]
\includegraphics[width=0.485\textwidth,keepaspectratio]{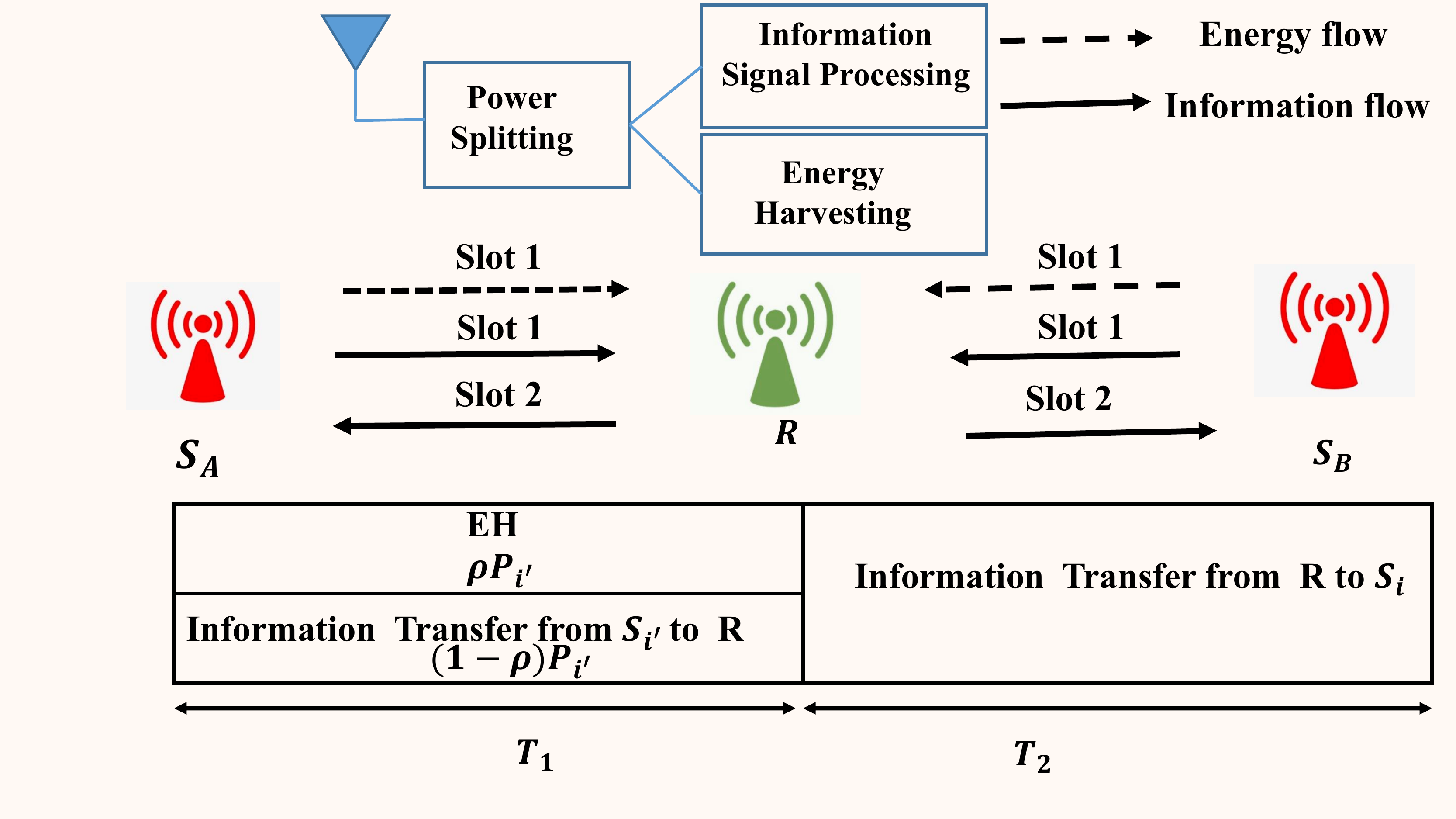}
\caption{ In the considered system model, sources $A$ $(S_A)$ and  $B$ $(S_B)$ exchange status updates with each other with the help of a single relay $(R)$; sources send updates to the $R$  during the first transmission time slot $T_{1}$  while the $R$ is harvesting energy and then the $R$ exchanges updates received from each source using harvested energy during the second transmission time slot $T_{2}$.}
\label{sytmmod}      
\end{figure}This paper considers a two-way cooperative status update system where two source nodes, source $A$ $(S_A)$ and source $B$ $(S_B)$ exchange status updates with each other as timely as possible with the help of a single relay $(R)$. $R$ adopts the decode-and-forward relaying protocol. Specifically, $S_{i},i\in \left ( A, B \right )$ transmits status updates which can be generated at the beginning of any time slot. The sources in this system are regarded as energy providers and the relay is equipped with an energy harvesting device and is capable of conducting data forwarding and harvesting energy simultaneously.  In this paper, we assume $R$ adopts the dynamic power splitting technique \cite{perera2017simultaneous}. As shown in \figurename{ \ref{sytmmod}}, we consider a two-time slot transmission scheme in which $S_{i}$ sends updates to the $R$ while harvesting energy during the first transmission time slot and then the $R$ exchange updates received from $S_{i}$ using harvested energy during the second transmission time slot. Specifically, $S_{i},i\in \left ( A, B \right )$ transmits status updates which can be generated at the beginning of any time circle following  generate-at-will updates generation model \cite{ceran2019average}. Under this policy, the relay uses energy harvested within the first transmission time slot ($T_{1}$) for the transmission in the second transmission time slot ($T_{2}$) without waiting. Suppose the harvested energy is less than the minimum required energy for the transmission. In that case, the relay does not transmit received updates and updates received from both sources are destroyed at the relay. $H_{ij}$ represents the channel coefficient of the channel between node $i$ to node $j$ where   $i,j\in \left \{ A,B,R \right \}$ and $i\neq j$. The small scale channel gain  is $g_{\textit{i}\textit{j}}=\left | h_{\textit{i}\textit{j}}\right |^2$, where $h_{\textit{i}\textit{j}} \sim  \mathcal{C}\mathcal{N}(0,1)$  is the Rayleigh fading channel coefficient. The probability density function (PDF) of the small scale  channel gain is defined as  $
f_{g_{\textit{i}\textit{j}}}(z)=e^{-z},z\geq 0$. The large scale channel gain $\alpha_{ij}$ is given by 
   $-10\log_{10}(\alpha_{\textit{i}\textit{j}} )= 20 \log_{10}(d_{\textit{i}\textit{j}})+20\log_{10}(\frac{4\pi f_{c}}{c})$, where $f_{c}, d_{\textit{i}\textit{j}}$ are the carrier frequency and distance between node $i$ and node $j$ respectively. $c$ is the speed of light in the space. Thus, channel coefficient is written  as $H_{ij}=\sqrt{\alpha_{ij}g_{ij}}$ and channel gain is written  as $ G_{ij}=\alpha_{ij}g_{ij}$.
Up-link transmission between the source and the relay are performed in an orthogonal channel. 
\section{Block error analysis under finite block-length }
Our first objective is to study   block error probability  at each destination (opposite source) in this system. The  block error probability  at the destination has been derived using different mathematical approximation techniques in this section.  \par To calculate block error probability  at each node, it is necessary to derive the SNR at each node.
The received SNR  at the relay from each source node $S_{{i}'},{i}' \in \left \{ A,B \right \}$ is given by, 
\begin{equation}
    \gamma^{{i}'}_{R}=\frac{(1-\rho) P_{{i}'}G_{{i}'R}}{\sigma_{R} ^{2} },
    \label{snratrfrs}
\end{equation}
where $\rho $ is the power spitting factor, noise power at the relay is denoted as $\sigma^{2}_R$ and transmit power at the $S_{{i}'}$ given by $P_{{i}'} $. Then, the energy harvested by the relay from the each node is given by $ E^{{i}'}_{R}= \rho \eta P_{{i}'}G_{{i}'R}T_{1}$, where  $\eta$\ is the energy conversion efficiency and $T_{1}$ is the transmission time of the first transmission slot  and it can be calculated as $T_{1}=n^{{i}'}_{R}T_{s}$, where $n^{{i}'}_{R}$ is the allocated block length for the transmission between ${i}'$ and $R$ and $T_{s}$ is the symbol duration. Then, the total energy harvested by the relay within the first transmission slot is given by  
\begin{equation}
    E_{R}=\sum_{{i}' = \left \{ A,B \right \}}^{} E^{{i}'}_{R}=\rho \eta T_{1}\left ( P_{A}G_{AR} +P_{B}G_{BR}\right ).
    \label{havestpower}
\end{equation}
The available energy harvested by the relay for transmission is given by \begin{equation}
 E_{R}^{T}=
\begin{cases}
E_{max}  &   E_{R}\geq   E_{max} \\ 
E_{R}  &  E_{max}> E_{R}> E_{min} \\ 
0 & \text{otherwise},
\end{cases}
\label{avaitransp}
\end{equation}where  $E_{max}$ is the maximum energy limit that the relay can harvest and $E_{min}$ is the minimum energy required for transmission.  $E_{min}$ can be calculated as 
$E_{min}=P_{min}T_{2}$, where  $P_{min}$ is the minimum required power for the transmission while $T_{2}$ is the  time of the second transmission slot and $T_{2}$ is calculated as $ T_{2}=\sum_{i = \left \{ A,B \right \}}^{} n^{R}_{i}T_{s}$, where $n^{R}_{i}$ is the allocated block length for the transmission between $R$ and $i$. Then, the transmit power of relay  is calculated as $P_{R}=\frac{E_{T}^{R}}{T_{2}}$.
During the second time slot, the received SNR at each  $S_{i} \in\left \{ A,B \right \}$ is given by $ \gamma_{i}=\frac{ P_{R}G_{Ri}}{\sigma_{i} ^{2} }$
Then, using  \eqref{havestpower} and \eqref{avaitransp} SNR at the destination is expressed as
\begin{equation}
     \gamma _{i}=
\begin{cases}
 \frac{E_{max}\alpha_{Ri}g_{Ri}}{T_{2}\sigma_{i} ^{2} }, &   E_{R}>   E_{max} \\ 
\frac{E_{R}\alpha_{Ri}g_{Ri}}{T_{2}\sigma_{i}^{2} }, &  E_{max}\geq  E_{R}\geq  E_{min} \\ 
0 ,& \text{otherwise}.
\end{cases}
\end{equation}
An outage happens  when the relay or opposite source are unable to decode the received message successfully. Hence, the system overall transmission success probability $\varphi_{i} $ at each source node $i$ can be calculated as
\begin{equation}
   \varphi_{i}= 1-\left ( \varepsilon^{{i}'}_{R}+\left (1-\varepsilon^{{i}'}_{R}  \right ) \varepsilon^{R}_{i} \right ),
   \label{sucessprob}
\end{equation}
where $i\neq {i}' , i,{i}' \in \left \{ A,B \right \}$ and $\varepsilon^{t}_{j}$ is the decoding error probability at receiving node  $j\in \left ( i,R \right )$ for block received from node $t\in \left ( {i}',R \right )$. Due to the static nature of the communication channels, it is assumed that the fading coefficients stay constant over the duration of each transmission block. Following Polyanskiy's results on short packet communication \cite{polyanskiy2010channel} and assuming that the receiver has the perfect channel state information, the expectation of the block error probability at the receiving node for a given block length $n^{t}_{j}$ can be written as
    \begin{equation}
        \varepsilon^{t}_{j}=\mathbb{E}\left [ Q\left (\frac{n^{t}_{j}C(\gamma^{t}_{j})-k^{t}_{j}}{\sqrt{n^{t}_{j}V(\gamma^{t}_{j})}} \right ) \right ],
    \end{equation}
where  $\mathbb{E}\left [ . \right ]$ is the expectation operator, $Q(x)=\frac{1}{\sqrt{2\pi }}\int_{x}^{\infty }e^{-\frac{t_{2}}{2}}dt$ and $V(\gamma^{t}_{j})$ is the  channel dispersion, which can be written $V(\gamma^{t}_{j} )=\frac{{\log _{2}}^{2}e}{2}(1-\frac{1}{(1+\gamma^{t}_{j} )^2})$. The variable $C(\gamma^{t}_{j})$ denotes the channel capacity of a complex AWGN channel and it is given by $C(\gamma^{t}_{j})=\log _{2}(1+\gamma^{t}_{j})$. The number of bits per block represents by $k^{t}_{j}$. Moreover, under the Rayleigh fading channel conditions, $\varepsilon^{t}_{j} $ can be formulated as
    \begin{equation}
      \varepsilon^{t}_{j} = \int_{0}^{\infty }f_{\gamma^{t}_{j}} (z)Q\left (\frac{n^{t}_{j}C(\gamma^{t}_{j})-k^{t}_{j}}{\sqrt{n^{t}_{j}V(\gamma^{t}_{j})}} \right)dz,
      \label{e11}
    \end{equation}
where  $f_{\gamma^{t}_{j}}(z)$ denotes the PDF of the received SNR ($\gamma^{t}_{j}$) at the receiving node $j$. Due to the complexity of the Q-function, it is challenging to get a closed-form expression for the overall decoding error probability. Thus, using the approximation technique given in \cite{makki2014finite}  and \cite{gu2017ultra}, \eqref{e11} can be approximated as $
        \varepsilon^{t}_{j} \approx  \int_{0}^{\infty }f_{\gamma^{t}_{j}} (z)\Theta^{t}_{j} (z) dz$,
where $\Theta^{t}_{j}(z)$ denotes the linear approximation of $Q\left ( \frac{n^{t}_{j}C(\gamma^{t}_{j})-k^{t}_{j}}{\sqrt{n^{t}_{j}V(\gamma^{t}_{j}})} \right)$, this can be expressed as in \cite{gu2017ultra}
    \begin{equation}
        \Theta^{t}_{j} (z)=\left \{ \begin{matrix}
1,&  \gamma^{t}_{j}\leq \phi^{t}_{j},  & \\ 
\frac{1}{2}-\beta^{t}_{j} \sqrt{n^{t}_{j}}(\gamma^{t}_{j}-\psi^{t}_{j}), & \phi^{t}_{j}< \gamma^{t}_{j} <\delta^{t}_{j}, & \\ 
 0,&  \gamma^{t}_{j} \geq \delta^{t}_{j},& 
\end{matrix} \right.
\label{qfap}
    \end{equation}
where $\beta^{t}_{j} =\frac{1}{2\pi \sqrt{2^{\frac{2k^{t}_{j}}{n^{t}_{j}}}-1}},\psi_{j}=2^{\frac{k^{t}_{j}}{n^{t}_{j}}}-1,\phi^{t}_{j}=\psi^{t}_{j}-\frac{1}{2\beta^{t}_{j}\sqrt{n^{t}_{j}}}$ and $\delta^{t}_{j}=\psi^{t}_{j}+\frac{1}{2.\beta^{t}_{j}\sqrt{n^{t}_{j}}}$. By using above linear approximation $\varepsilon^{t}_{j}$ can be expressed as
\begin{equation}
   \varepsilon^{t}_{j} \approx \beta^{t}_{j}\sqrt{n^{t}_{j}}\int_{\phi^{t}_{j}}^{\delta^{t}_{j}}F_{\gamma^{t}_{j}}\left ( z \right )dz,
   \label{aprxberwcdf}
\end{equation} where $F_{\gamma^{t} _{j}}\left ( z \right )$ denotes the CDF of the received SNR ($\gamma^{t}_{j}$) at receiving node $j$. To calculate success probability at each source $i$ using \eqref{sucessprob}, it is necessary to calculate $\varepsilon^{{i}'}_R$ and $\varepsilon^{R}_{i}$. Using (\ref{qfap}) and (\ref{aprxberwcdf}) block error probabilities at the $R$ and each source can be calculated as follows,
\begin{equation}
   \varepsilon^{{i}'}_{R} \approx \beta^{{i}'}_{R}\sqrt{n^{{i}'}_{R}}\int_{\phi^{{i}'}_{R}}^{\delta^{{i}'}_{R}}F_{\gamma^{{i}'}_{R}}\left ( z \right )dz,
      \label{aprxberwcdf_R}
\end{equation}
\begin{equation}
   \varepsilon^{R}_{i} \approx \beta^{R}_{i}\sqrt{n^{R}_{i}}\int_{\phi^{R}_{i}}^{\delta^{R}_{i}}F_{\gamma^{R}_{i}}\left ( z \right )dz.
\end{equation}
\begin{lemma}
An approximation for the block error probability at the relay is derived as
\begin{multline}
\varepsilon^{{i}'}_{R}  \approx 1-\left (\frac{(1-\rho) P_{{i}'}\alpha _{{i}'R}\beta^{{i}'}_{R}\sqrt{n^{{i}'}_{R}}}{\sigma_{R} ^{2}}  \right ) \\
\left ( e^{-\frac{\phi^{{i}'}_{R}\sigma_{R} ^{2}}{(1-\rho) P_{{i}'}\alpha _{{i}'R}}}-e^{-\frac{\delta^{{i}'}_{R}\sigma_{R} ^{2}}{(1-\rho) P_{{i}'}\alpha _{{i}'R}}} 
 \right ).
 \label{policy1errsr}
\end{multline}
    \begin{proof}
The PDF of SNR at relay from each source can be calculated using \eqref{snratrfrs} as
$
f_{\gamma^{{i}'}_{R}}{}\left ( x \right )=\frac{\sigma_{R} ^{2}}{(1-\rho) P_{{i}'}\alpha _{{i}'R}}e^{-\frac{x\sigma_{R} ^{2}}{(1-\rho) P_{{i}'}\alpha _{{i}'R}}}$ Then, the CDF can be calculated as
\begin{equation}
\begin{aligned}
F_{\gamma^{{i}'}_{R}}\left ( z \right )&= 1-e^{-\frac{z\sigma_{R} ^{2}}{(1-\rho) P_{{i}'}\alpha _{{i}'R}}}.
\label{cdfsnratrvt}
\end{aligned}
\end{equation}
Then, the result can be proved by substituting \eqref{cdfsnratrvt} to $\eqref{aprxberwcdf_R}$.
\end{proof}
\end{lemma}
 \begin{lemma}
Block  error probability at the opposite receiving node can be derived  as follows:
\begin{equation}
  \varepsilon^{R}_{i}\approx  \beta^{R}_{i}\sqrt{n^{R}_{i}}\left ( \left ( \frac{\delta ^{R}_{i}+\phi^{R}_{i}}{2} \right )\sum_{v=1}^{V}\frac{\pi }{V}\sqrt{1-\phi_{v}^2}F_{\gamma^{R}_{i}}\left ( q \right )+R_{V} \right )
   \label{policy1er}
\end{equation}
where  $\phi_{v}=\cos \left ( \frac{2v-1}{2v} \pi \right )$, $q=\left ( \frac{\delta^{R}_{i}-\phi^{R}_{i}}{2} \right )\phi_{v}+\left ( \frac{\delta^{R}_{i}+\phi^{R}_{i}}{2} \right )$, $V$ is the complexity-accuracy trade-off factor, while $R_{V}$ denotes the error term, which is ignored at substantially larger values of $V$.
\begin{proof}
To calculate error probability at  $S_{i}$, it is necessary to derived the CDF of SNR. Using \eqref{avaitransp}, $F_{\gamma^{R}_{i}}\left ( z \right )$ is evaluated as
\begin{equation}
   \begin{aligned}
    F_{\gamma^{R}_{i}}\left ( z \right )= &\mathbb{P}_{r}\left ( \gamma^{R}_{i}< z \right )= 1- \mathbb{P}_{r}\left \{ E_{R} \geq E_{min} \cap \gamma^{R}_{i}> z\right \}\\
F_{\gamma^{R}_{i}}\left ( z \right )=& 1- \underbrace{\mathbb{P}_{r} \left \{  E_{min}\leq E_{R}\leq E_{max}\cap\gamma^{R}_{i}> z\right \}}_{L_{1}} \\ & - \underbrace{\mathbb{P}_{r} \left \{  E_{R}\geq  E_{max}\cap\gamma^{R}_{i}> z\right \}}_{L_{2}}.
   \end{aligned}
   \label{apen1}
\end{equation}
Then, substituting $E_{R}=\rho \eta T_{1}\left ( P_{A}\alpha _{AR}g_{AR} +P_{B}\alpha_{BR}g_{BR}\right )$ into \eqref{apen1}, $L_{1}$ is evaluated as
\begin{equation}
\begin{aligned}
    L_{1}= &\mathbb{P}_{r}\left \{  \Omega_{1} < I<  \Omega _{2} \, \cap \, Ig_{Ri}> \Omega _{3} \right \},
       \end{aligned}
\end{equation}
or
\begin{equation}
    L_{ 1}=
\begin{cases}
0, & g_{Ri}< \frac{\Omega _{3}}{ \Omega_{2}}, \\
\mathbb{P}_{r}\left \{ \frac{\Omega_{3}}{g_{Ri}}< I\leqslant \Omega _{2}\right \},&\frac{\Omega _{3}}{\Omega  _{2}} <g_{Ri}<\frac{\Omega _{3}}{\Omega _{1}}, \\
\mathbb{P}_{r}\left \{ \Omega_{1}< I\leqslant \Omega _{2}\right \},& g_{Ri}> \frac{\Omega  _{3}}{\Omega _{1}},
\end{cases}
\end{equation}
where $\Omega  _{1}=\frac{E_{min}}{\rho \eta T_{1}}$ , $\Omega _{2}=\frac{E_{max}}{\rho \eta T_{1}}$, $\Omega _{3}=\frac{z\sigma _{i}^{2}T_{2}}{\rho \eta T_{1}\alpha _{Ri}}$ and $ I=\sum_{i=\left \{ A,B \right \}}^{}  P_{i}\alpha _{iR}g_{iR}$. To calculate $L_{1}$ it is necessary to get PDF and CDF of $I$ and $g_{Ri}$. Then, to calculate the PDF of $I$, it is considered as summation of two independent random variable as $I=\mu_{1}+\mu_{2 }$, where $\mu_{1}\sim \mathrm{exp}(\frac{1}{P_{A}\alpha_{AR}})$ and $\mu_{2} \sim \mathrm{exp}(\frac{1}{P_{B}\alpha_{BR}})$. Using the concepts of convolution of random variables, PDF and CDF of $I$ can be calculated as follows:
\begin{equation}
\begin{aligned}
f_{I}\left ( z \right ) & =\int_{-\infty }^{\infty  } f_{\mu_{1}}(x)f_{\mu_{2}}(z-x)dx,\\ 
 & = \int_{0}^{z}\frac{1}{P_{A}\alpha_{AR}}e^{-\frac{1}{P_{A}\alpha_{AR}}x}\frac{1}{P_{B}\alpha_{BR}}e^{-\frac{1}{P_{B}\alpha_{BR}}(z-x)}dx,\\ 
 &  =
\frac{1}{P_{A}\alpha_{AR}P_{B}\alpha_{BR}}e^{-\frac{1}{P_{B}\alpha_{BR}}z}\int_{0}^{z}e^{(\frac{1}{P_{B}\alpha_{BR}}-\frac{1}{P_{A}\alpha_{AR}})x}dx,\\ 
f_{I}\left ( z \right ) & =
\begin{cases}
\frac{1}{P_
{A}\alpha_{AR}-P_{B}\alpha_{BR}}(e^{-\frac{1}{P_{A}\alpha_{AR}}z}-e^{-\frac{1}{P_{B}\alpha_{BR}}z}), \\ \qquad \textrm{if} \quad \frac{1}{P_{A}\alpha_{AR}}\neq \frac{1}{P_{B}\alpha_{BR}},\\ 
\frac{1}{(P\alpha)^2}ze^ {-\frac{1}{P\alpha}z}, \quad \textrm{if}\, \:  \frac{1}{P_{A}\alpha_{AR}}= \frac{1}{P_{B}\alpha_{BR}}=\frac{1}{P\alpha},
\end{cases}.
\end{aligned}
\end{equation} where $f$ denotes the PDF function of a random variable.
Then, the CDF of $I$ can be calculated as
\begin{equation}
   \begin{aligned}
F_{I}(z)&=P(Z\leq z)=\int_{0}^{z} f(t) dt, \\
F_{I}(z)&=\begin{cases}
 1+\frac{P_{B}\alpha_{BR}}{P_A \alpha_{AR}-P_{B}\alpha_{BR}}e^{-\frac{z}{P_{B}\alpha_{BR}}}\\ \quad -
 \frac{P_{A}\alpha_{AR}}{P_{A}\alpha_{AR}-P_{B}\alpha_{BR}}e^{-\frac{z}{P_{A}\alpha_{AR}}}, & \\ ,\textrm{if}\,  P_{A}\alpha_{AR}\neq P_{B}\alpha_{BR} ,& \\ 
 1-e^{-\frac{1}{P\alpha}z}\left ( 1+\frac{1}{P\alpha}z \right ) \textrm{if}\, \:  \frac{1}{P_{A}\alpha_{AR}}= \frac{1}{P_{B}\alpha_{BR}}=\frac{1}{P\alpha},
\end{cases}
\end{aligned}
\end{equation}
Further, we derive the approximation for $L_{1}$ as follows,
\begin{equation}
        \begin{aligned}
    L_{1} &=\int_{\frac{\Omega_{3}}{\Omega_{2}}}^{\frac{\Omega_{3}}{\Omega_{1}}}\left ( f_{g_{Ri}}\left ( x \right )\int_{\frac{\Omega_{3}}{x}}^{\Omega _{2}} f_{I}\left (y  \right )dy\right )dx\\ &  \:  \: \: +\int_{\Omega_{1}}^{\Omega_{2}}f_{I}\left ( x \right )dx\int_{\frac{\Omega _{3}}{\Omega _{1}}}^{\infty }f_{g_{Ri}}\left ( x \right )dx,\\
       L_{1} =& L_{3}+ \left ( F_{I}\left ( \Omega_{2} \right ) -F_{I}\left ( \Omega_{1} \right )\right )\left ( 1-F_{Ri}\left ( \frac{\Omega _{3}}{\Omega_{1}} \right ) \right ),
    \end{aligned}
\end{equation}
where 
\begin{equation}
 \begin{aligned}
L_{3}=&\int_{\frac{\Omega _{3}}{\Omega _{2}}}^{\frac{\Omega _{3}}{\Omega _{1}}}\left ( f_{g_{Ri}}\left ( x \right )\int_{\frac{\Omega _{3}}{x}}^{\Omega_{2}} f_{I}\left (y  \right )dy\right )dx,
\\ =&\int_{\frac{\Omega _{3}}{\Omega_{2}}}^{\frac{\Omega _{3}}{\Omega _{1}}} f_{g_{Ri}}\left ( x \right )\left [ F_{I}\left ( \Omega _{2} \right )-F_{I}\left ( \frac{\Omega _{3}}{x} \right ) \right ]dx,\\
L_{3}=& F_{I}\left ( \Omega _{2} \right )\left [ F_{g_{Ri}}\left (  \frac{\Omega _{3}}{\Omega _{1}}\right ) -F_{g_{Ri}}\left (  \frac{\Omega _{3}}{\Omega _{2}}\right )\right ]-L_{4},
 \end{aligned}
\end{equation}
where $L_{4}$ is defined as
\begin{equation}
    \begin{aligned}
L_{4}=&\int_{\frac{\Omega _{3}}{\Omega _{2}}}^{\frac{\Omega _{3}}{\Omega_{1}}}f_{g_{Ri}}\left ( x \right )F_{I}\left ( \frac{\Omega_{3}}{x} \right )dx,
\label{erl4}
 \end{aligned}
\end{equation}
Using Gaussian-Chebyshev-Quadrature (GCQ) method \cite{abramowitz1988handbook}, \eqref{erl4} can be approximated as follows,
\begin{equation}
    L_{4}\approx \frac{\frac{\Omega_{3}}{\Omega _{1}}+\frac{\Omega _{3}}{\Omega _{2}}}{2}\sum_{m=1}^{M}\frac{\pi }{M}\sqrt{1-\phi _{m}^2}f_{g_{Ri}}\left ( z_{1} \right )F_{I}\left ( \frac{\Omega_{3}}{z_{1}} \right )+R_{M},
\end{equation}
where $\phi _{m}=\cos \left ( \frac{2m-1}{2M}\pi \right )$, $z_{1}=\frac{\frac{\Omega _{3}}{\Omega _{1}}-\frac{\Omega _{3}}{\Omega _{2}}}{2}\phi _{m}+\frac{\frac{\Omega _{3}}{\Omega _{1}}+\frac{\Omega _{3}}{\Omega _{2}}}{2}$, $M$ is the complexity-accuracy trade-off factor, and $R_{M}$ is the error term that can be ignored at sufficiently high $M$ values. Finally, expression for $L_{1}$ can be approximated as shown in \eqref{longl1}. 
\begin{figure*}[!t]
% ensure that we have normalsize text
\normalsize
%\begin{dmath}
\begin{multline}
  L_{1}\approx F_{I}\left ( \Omega_{2} \right )\left [ F_{g_{Ri}}\left (  \frac{\Omega _{3}}{\Omega _{1}}\right ) -F_{g_{Ri}}\left (  \frac{\Omega_{3}}{\Omega _{2}}\right )\right ]-\frac{\frac{\Omega _{3}}{\Omega_{1}}+\frac{\Omega_{3}}{\Omega _{2}}}{2}\sum_{m=1}^{M}\sqrt{1-\phi _{m}^2}f_{g_{Ri}}\left ( z_{1} \right )F_{I}\left ( \frac{\Omega_{3}}{z_{1}} \right )+R_{M}+ \\
  \left ( F_{I}\left ( \Omega_{2} \right ) - F_{I}\left ( \Omega_{1} \right )\right )\left ( 1-F_{g_{Ri}}\left ( \frac{\Omega _{3}}{\Omega_{1}} \right ) \right ).
    \label{longl1}
\end{multline}
\begin{multline}
F_{\gamma^{R}_{i}}\left ( z \right )  \approx 1- F_{I}\left ( \Omega _{2} \right )\left [ F_{g_{Ri}}\left (  \frac{\Omega  _{3}}{\Omega _{1}}\right ) -F_{g_{Ri}}\left (  \frac{\Omega _{3}}{\Omega _{2}}\right )\right ]-\frac{\frac{\Omega _{3}}{\Omega_{1}}+\frac{\Omega_{3}}{\Omega _{2}}}{2}\sum_{m=1}^{M}\sqrt{1-\phi _{m}^2}f_{g_{Ri}}\left ( z_{1} \right )F_{I}\left ( \frac{\Omega_{3}}{z_{1}} \right )+R_{M}+ \\
  \left ( F_{I}\left ( \Omega_{2} \right ) - F_{I}\left ( \Omega_{1} \right )\right ) \left ( 1-F_{g_{Ri}}\left ( \frac{\Omega _{3}}{\Omega_{1}} \right ) \right ) -\left ( 1-F_{I} \left (\Omega _{2} \right )\right ) \left (1-F_{g_{Ri}} \left ( \Omega_{4} \right ) \right ).
  \label{snrapatdes}
\end{multline}
\hrulefill
\vspace*{4pt}
\end{figure*}
Similarly, $L_{2}$ is calculated as
\begin{equation}
 \begin{aligned}
   L_{2}&= \mathbb{P}_{r} \left \{  I>   \Omega_{2}\cap g_{Ri}> \Omega _{4}\right \}\\
   &=\left ( 1-F_{I} \left ( \Omega _{2} \right )\right )\left (1-F_{g_{Ri}} \left ( \Omega _{4} \right ) \right ),
   \label{l1eq}
   \end{aligned}
\end{equation}
where $\Omega  _{4}= \frac{z\sigma_{i} ^{2}T_{2}}{E_{max}\alpha _{Ri}}$. Then, CDF of SNR  at each  destination ($F_{\gamma^{R}_{i}}\left ( z \right ) $) can be obtained by substituting \eqref{longl1} and \eqref{l1eq} in \eqref{apen1} as in \eqref{snrapatdes}. Then, the result can be proved by substituting \eqref{snrapatdes} to \eqref{aprxberwcdf} and then applying the GCQ method for the integration of the CDF function.
\end{proof}
\end{lemma} Finally, substituting \eqref{policy1errsr} and \eqref{policy1er} into \eqref{sucessprob} the overall transmission success probability can can be calculated.
\section{Age of Information Analysis}
This section estimates the AAoI of the two-way relay system. This system adopts the generate-at-will update generation model \cite{ceran2019average}. Hence, $S_{A}$ and $S_{B}$ generate new status updates every transmission cycle to keep the information at the corresponding destinations as fresh as possible. Then, generated updates are transmitted to their opposite sources using a relay system. If the generation time of the freshest update received at opposite source  time stamp $t$ is $g(t)$, then AoI can be defined as a random process as $
 \Delta \left ( t \right )=t-g(t).$
\begin{figure}[!htb]
\includegraphics[width=0.5\textwidth]{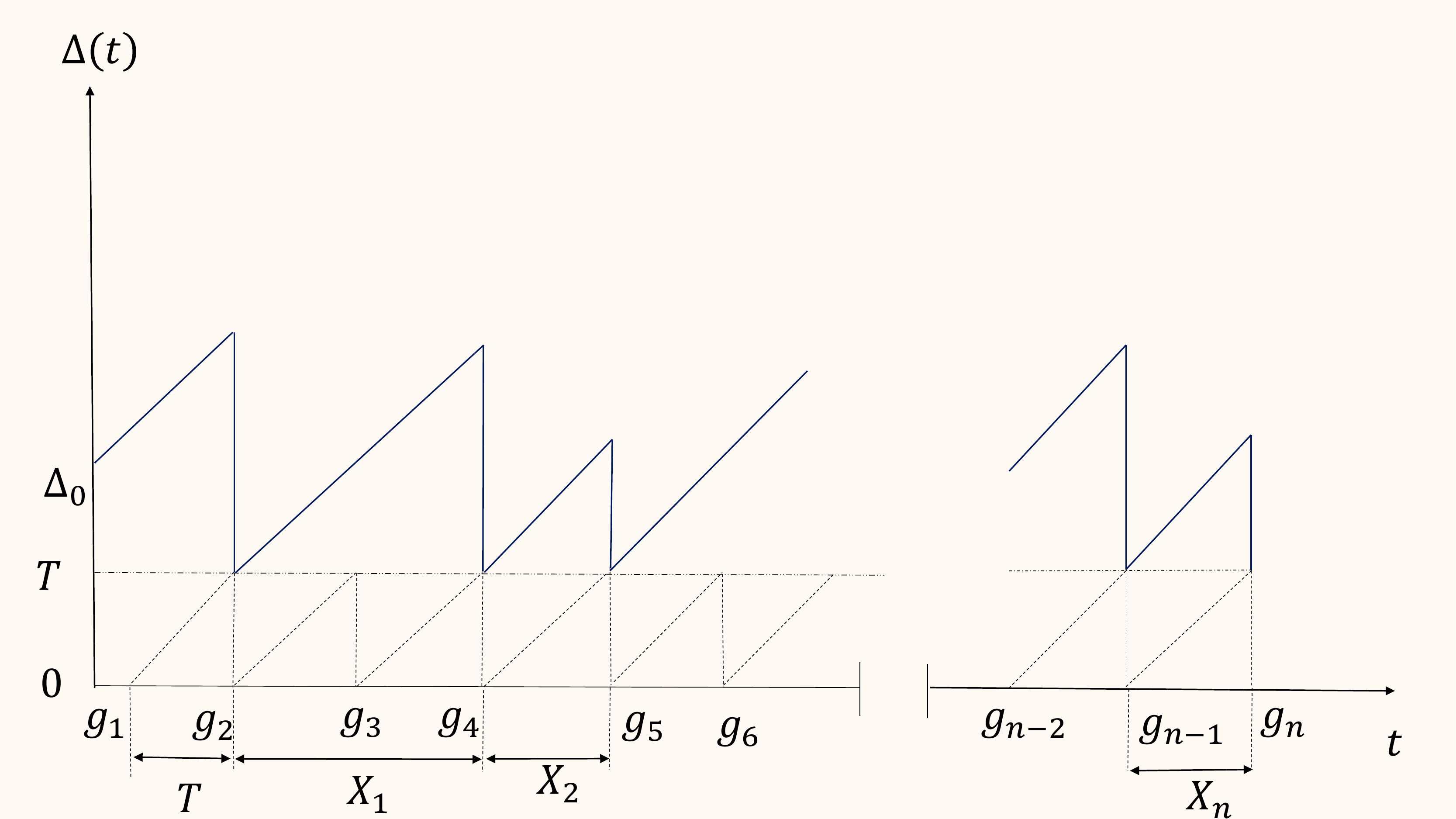}
\caption{ Evolution of AoI $\Delta(t)$ with the time: Each source generate updates at time stamps $\mathit{g_{1},g_{2},...,g_{n-1}}$ and the opposite source receive these updates at time stamps $\mathit{g_{2},g_{3},...,g_{n}}$; $\Delta(t)$ is the AoI at the opposite source (destination).}
\label{f2a}      
\end{figure}
As illustrated in the \figurename {\ref{f2a}}, it is assumed that at $t=0$ the measurements of the AoI starts and the AoI at the opposite source (destination) is set to  $\Delta (0)=\Delta  _{0}$. Each source generates updates at time stamps $\mathit{g_{1},g_{2},...,g_{n-1}}$ and the opposite source receive these updates at time stamps $\mathit{g_{2},g_{3},...,g_{n}}$. As illustrated in \figurename {\ref{f2a}}, data update $i$ is transmitted from the source at time stamp $t = g_{i}$ and it is successfully delivered to its opposite source  at time stamp $g_{i+1}= g_{i}+T$ where $T$ is total time allocated for a one transmission circle and $T=T_{1} +T_{2}$. Therefore, if update packet delivered successfully, at the time $g_{i+1}$, the AoI at the opposite source is estimated as $\Delta  (g_{i+1})=T$. We assume that AoI increases linearly until the next update is successfully delivered to the opposite source. As an example, one packet fails to be decoded at time $g_{3}$, hence, $\Delta(t)$ continues to increase linearly. For the considered time period $T_{c}$, time average AoI can be computed using the area under $\Delta(t)$. Similarly, the time average age  is estimated  as $\Delta_{{T_{c}}}= \frac{1}{T_{c}} \int_{0}^{T_{c}} \Delta(t)dt.$
 Similar to the work presented in  \cite{kosta2017age}, the time average age ($\Delta_{{T_{c}}}$) tends to ensemble average age when  $T_{c}\rightarrow \infty $, i.e., which can be expressed as
 \begin{equation}
  \Delta^{AAOI}=\mathbb{E}\left [ \Delta \right ]=\lim_{t\rightarrow \infty }\mathbb{E}\left [ \Delta \left ( t \right )\right ]=\lim_{T_{c}\rightarrow \infty }\Delta_{{T_{c}}}.
      \label{aaoie}  
       \end{equation}
Applying graphical methods to saw-tooth age waveform in \figurename {\ref{f2a}}  and using [\citeonline{basnayaka2021agej}, eq.8] we can calculated AAoI at each $S_{i},i\in \left ( A,B \right )$ as follows:
\begin{equation}
    \Delta _{i}^{AAOI} =\frac{E[X_{i}^{2}]}{2E[X_{i}]} +T,
    \label{aaoif}
\end{equation} where $X_{i}$ denote the inter departure time between two consecutive successfully received status updates at $S_{i}$. It assumes that the end-to-end delay of each successfully received update is always a constant, which is given by $E[Y_{i}]=T_{1}+T_{2}=T$. The inter departure time $X_{i}$ is a geometric random variable with mean $E[X_{i}] = \frac{T}{\varphi_{i}}$ and second
moment $E\left [ X_{i}^2 \right ]=\frac{T^2\left ( 2-\varphi_{i}  \right )}{\varphi_{i}^2}$. 
\begin{lemma} For the two way relay network,  the  expression of the AAoI at each source can be obtained as follows:
\begin{equation}
    \Delta _{i}^{AAOI}= \frac{T}{2}+\frac{T}{\varphi_{i}}.
\end{equation}
\begin{proof}
 The result can be proved by substituting $E[X_{i}] = \frac{T}{\varphi_{i}}$ and  $E\left [ X_{i}^2 \right ]=\frac{T^2\left ( 2-\varphi_{i}  \right )}{\varphi_{i}^2}$ into (\ref{aaoif}).
\end{proof}
\end{lemma}
The expected weighted sum AAoI of the two-way relay system can  be calculated  as follows,
$
 \Delta _{Sum}^{AAoI}=\sum_{i=\left \{ A,B \right \}}^{}\omega _{i} \Delta _{i}^{AAOI}$,
where $\omega _{i}$ is the  weighting coefficient at $S_{i}$.
\section{Simulation results and discussions}
In this section, we present the analytical and numerical simulation results. Unless otherwise stated, the simulation parameters are set as: $d_{AR}$ $=$  \SI{30}{\metre}, $d_{BR}$ $=$ \SI{30}{\metre}, $f_{c}$ $=$ \SI{900}{\mega \hertz}, speed of the light (m/s) $=$ $3\times 10^{8} \SI{}{\metre}\SI{}{\second}^{-1} $, ${P}_{A}$ $=$ 1 W,  ${P}_{B}$ $=$ 1 W, $T_{s}$ $=$ \SI{20}{\micro \second}, $n^{A}_{R},n^{B}_{R}$ $=$ 200 bits, $n^{R}_{A},n^{R}_{B}$ $=$ 200 bits, $k^{A}_{R},k^{B}_{R}$ $=$ 32 bits, noise power ($\sigma_{R}^2,\sigma_{A}^2,\sigma_{B}^2$) $=$ -100 dBm,
, $E_{max}$ $=$ 0.001 J, $P_{min}$ $=$ 0.0001 mW, $\rho $ $=$ 0.5, $\omega _{A}$ $=$ 0.5, $\omega _{B}$ $=$ 0.5 and energy $\eta$ $=$ 0.9.
\begin{figure}[!htbp]
  \includegraphics[width= \linewidth  ,height=0.85\linewidth]{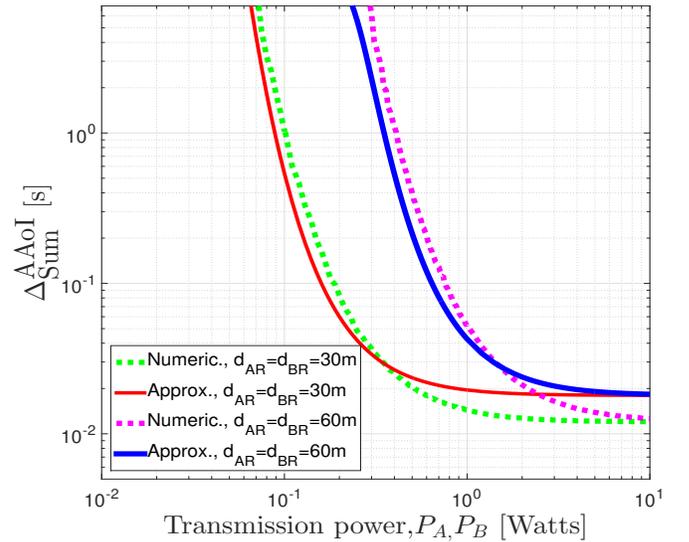}
\caption{ Weighed sum AAoI as a function of transmission power}
\label{sim1}      
\end{figure} 
\par The weighted sum of AAoI as a function of transmission power in Fig. \ref{sim1} for different distances is derived. The weighted sum of AAoI decreases dramatically as the transmit power at the sources increases, since increasing the transmit power at the source reduces the error probability at the relay node and increases the amount of energy harvested by the relay. However, for large transmission power levels, the AAoI value is fixed since the number of erroneous packets that impact the AAoI is too small. On the other hand, when the distance between the relay and sources is short, the AAoI is low, and when the distance increases, the AAoI increases due to low SNR. The numerically simulated AAoI well coincides with the approximated results, especially moderate SNR values, since the linear approximation applied in \eqref{qfap} is too tight for moderate SNR values \cite{basnayaka2021age}.
\begin{figure}[!htbp]
  \includegraphics[width= \linewidth  ,height=0.8\linewidth]{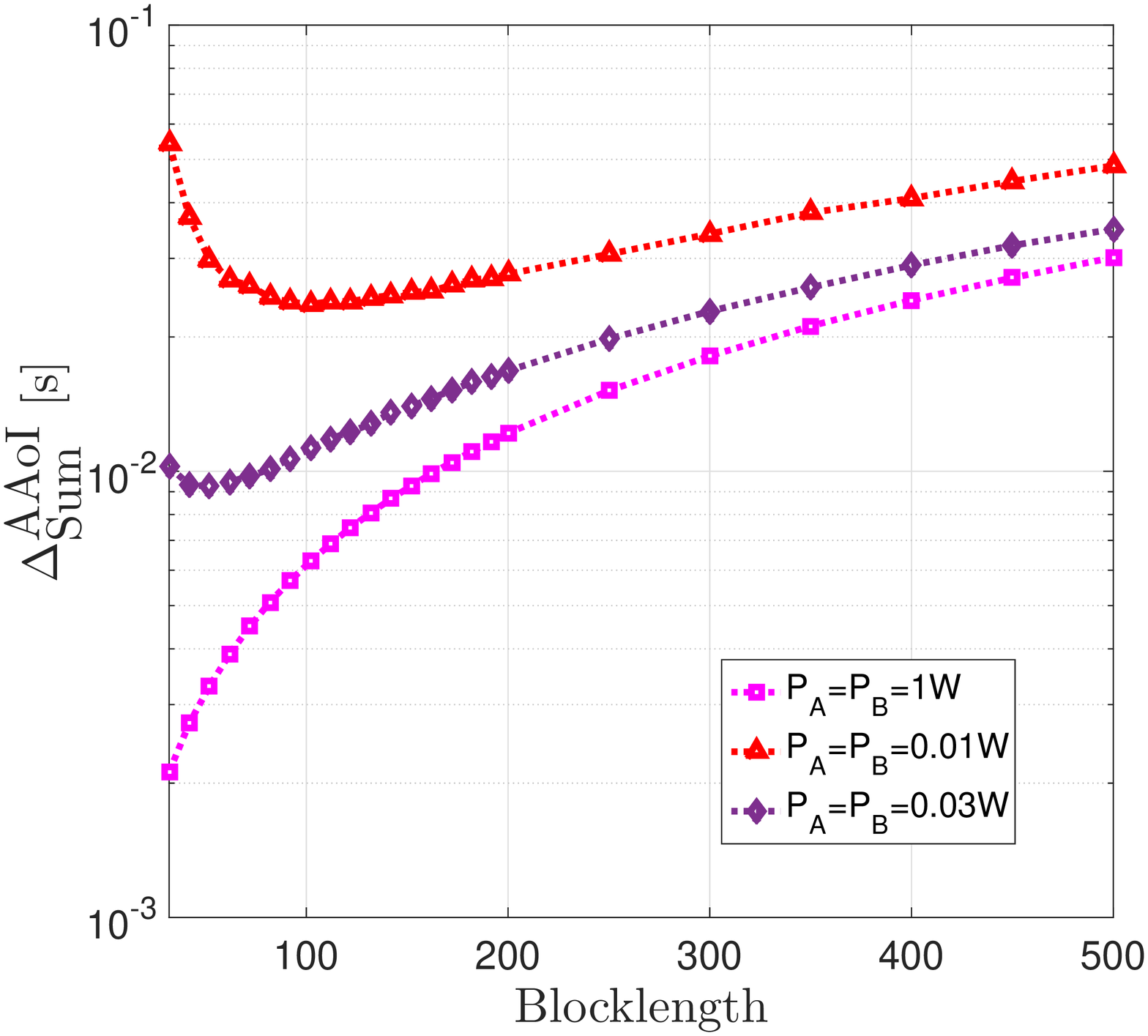}
\caption{ Weighed sum AAoI as a function of block length.}
\label{sim2}      
\end{figure} Next, in Fig. \ref{sim2}, we plot the weighted sum AAoI versus block length. When the transmission power is high, AAoI increases when block length is increased, as the number of erroneous packets is too low under high SNR conditions and increasing block length only increases transmission time. However, under low-SNR scenarios, small block length increases AAoI due to the high block error probability and increasing block length towards its optimal value decreases the AAoI due to the decrease in error probability. On the other hand, increasing the block length after the optimal value has resulted in an increase in AAoI because the impact of transmission time on AAoI is greater than the decrease in block error probability. This result proved that a short block length does not always maintain information freshness. 
\begin{figure}[!htbp]
  \includegraphics[width= \linewidth  ,height=0.8\linewidth]{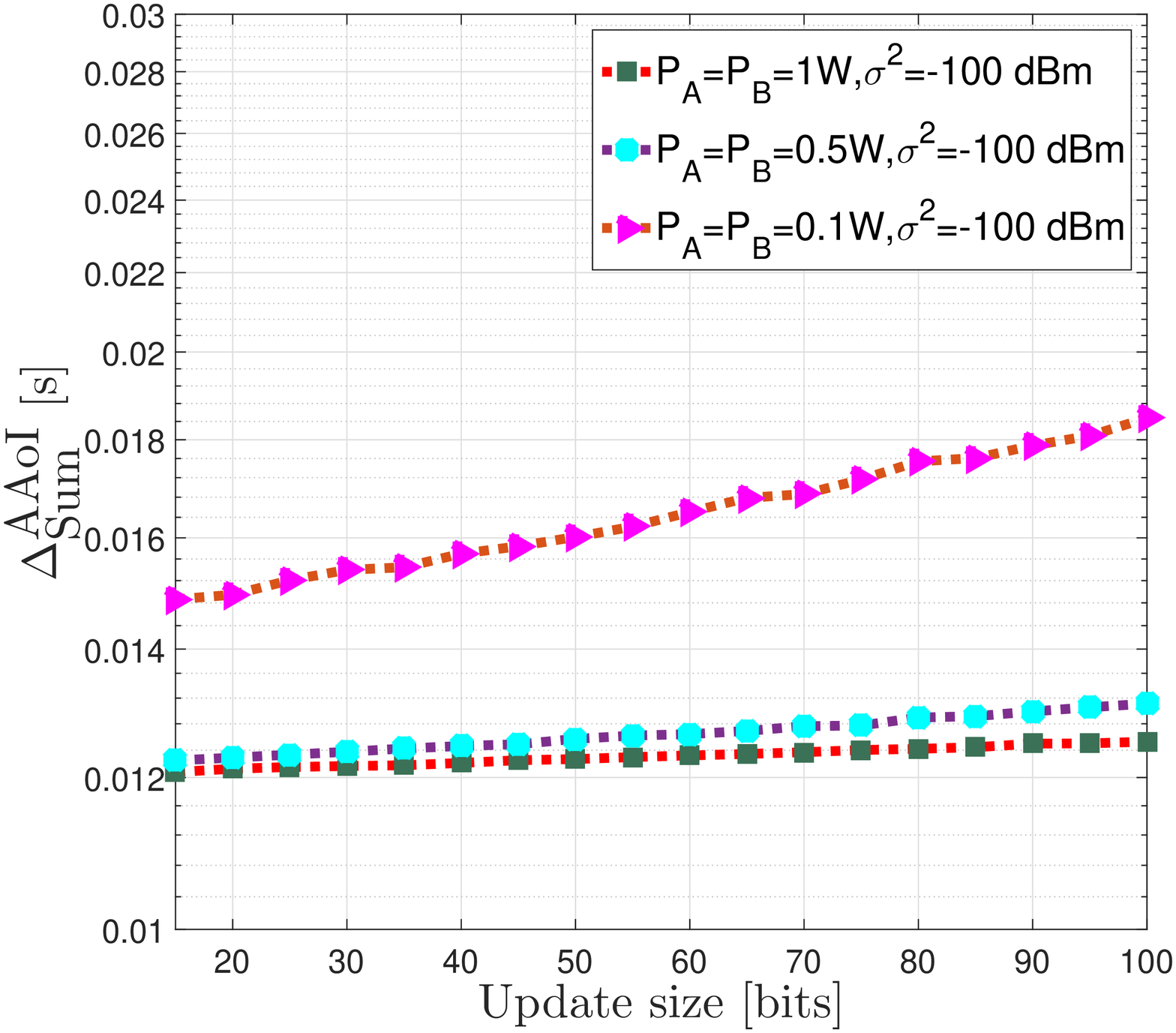}
\caption{ Weighed sum AAoI as a function of update size.}
\label{sim3}      
\end{figure}
In Fig. \ref{sim3} we present the weighed sum AAoI versus update size. If the transmission power is low, the AAoI increases as the packet size increases under fixed block length since it increases the overall block error probability. However, in high SNR scenarios, packet size has little effect on AAoI since block error probability is low.
\begin{figure}[!htbp]
\centering
  \includegraphics[width=0.48\textwidth,height=0.8\linewidth]{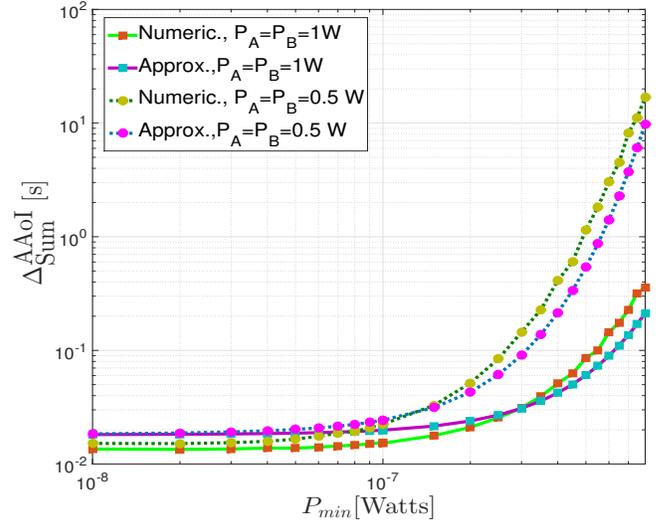}
\caption{Weighed sum AAoI as a function of $P_{min}$.}
\label{sim7}      
\end{figure}
\figurename{ \ref{sim7}} illustrates the weighted sum AAoI as a function  of $P_{min}$. The weighted sum AAoI of the system is monotonically increasing with the $P_{min}$ since high $P_{min}$ threshold values increase update loss at the relay.
\section{Conclusions} 
This work developed a model to estimate the AAoI in a two-way relay equipped with SWIPT that operates under ultra-reliable and low latency constraints. We derived an approximation for AAoI at each source using linear approximation techniques. The impacts of various parameters was studied,i.e., including block length, packet size, transmission power and noise level. Then, the numerical analysis to evaluate and validate the derived results. We observed that packet size does not affect freshness when SNR is high. Short packet communication retains an improved   AoI performance  in low SNR scenarios. This paper concludes that the  short block length communications does not always assist in maintaining freshness in SWIPT--enabled communications systems, even though it always assists in maintaining a low latency.

\section{Acknowledgement} 
This work is funded by the CEU-Cooperativa de Ensino Universit\'{a}rio, Portugal.
\bibliographystyle{IEEEtran}
    \bibliography{name}
\end{document}